\documentclass[
reprint,amsmath,amssymb,aps]{revtex4-2}

\usepackage{graphicx} 
\usepackage{float}
\usepackage[english]{babel}
\usepackage{array}
\usepackage{charter}
\usepackage[usenames,dvipsnames]{xcolor}
\usepackage{hyperref}
\usepackage{amsmath,amssymb,scalerel}  
\usepackage{bm,hyperref,mathtools}

\newcommand{\hf}{\hat{F}_0}
\newcommand{\hv}{\hat{\mathbf{v}}}
\newcommand{\hdr}{\hat{D}_R}
\newcommand{\htt}{\hat{t}}
\newcommand{\hxi}{\hat{\boldsymbol{\xi}}}
\newcommand{\hN}{\hat{\mathbf{N}}}
\newcommand{\unity}{\boldsymbol{1}}
\newcommand{\ud}{\textrm{d}}
\newcommand{\ue}{\textrm{e}}
\newcommand{\kbT}{k_\mathrm{B}T}
\newcommand{\tlj}{t_\mathrm{LJ}}

\definecolor{cream}{RGB}{222,217,201}

\begin{document}

\title{Passive particle in an active bath: how can we tell it is out of equilibrium?}
 \author{Jeanine Shea$^1$, Gerhard Jung, Friederike Schmid$^1$}
 \affiliation{%
 $^1$Institut f\"{u}r Physik, Johannes Gutenberg-Universit\"{a}t Mainz, 55099 Mainz, Germany.
 }%
 
 \begin{abstract} 
We study a passive probe immersed in a fluid of active particles.
Despite the system’s non-equilibrium nature, the trajectory of the probe
does not exhibit
non-equilibrium signatures: its velocity distribution remains Gaussian,
the second fluctuation dissipation theorem is not fundamentally
violated, and the motion does not indicate breaking of time reversal
symmetry. To tell that the probe is out of equilibrium requires
examination of its behavior \emph{in tandem} with that of the active
fluid: the kinetic temperature of the probe does not equilibrate to that
of the surrounding active particles.
As a strategy to diagnose non-equilibrium from probe trajectories alone,
we propose to examine their response to a small perturbation which
reveals a non-equilibrium signature through a violation of the first
fluctuation dissipation theorem.
\end{abstract}
 
 \maketitle

\section{Introduction}

Active systems exhibit vastly different behavior from passive systems.
When active and passive elements are mixed, the dynamics of the
passive elements in the system are also significantly altered. Since
the experiments of Wu and Libchaber~\cite{wu} reported enhanced
diffusion of a probe in a bacterial bath, a plethora of both
experimental and theoretical studies have been published on the
behavior of a probe in an active
bath~\cite{steff,Ash,Brady,Brady_curved,Leptos,Foffano,mino_2,suma1,maggi_fdt,maggi,Chaki,Maes_probeeq,omer,From_RohanJ}.
Nevertheless, in spite of an increasing volume of literature, a
comprehensive understanding of this system is still lacking --
especially regarding the question of how to use such a probe for
measuring bath properties. Recent studies of this system have derived
equations for the net force exerted on bodies immersed in active
baths, as well as the density profile surrounding such a
body~\cite{Ash,Brady,Brady_curved}. They reached the conclusion that
the net force on a symmetric probe in an active bath is zero, as one
would expect from symmetry. However, this begs the question: how can
we tell that a symmetric probe in an active bath is out of
equilibrium?

This problem is addressed in the present work. We consider a system of
a passive probe particle immersed in a bath of active Langevin
particles (ALPs).  Similar to active Brownian particles
(ABPs)~\cite{Stenhammar_MIPS, Redner_MIPS}, interacting ALPs with high
Peclet numbers undergo motility induced phase separation
(MIPS)~\cite{MITD,ABP_ALP} in sufficiently dense systems. However, in
the present investigation, we limit ourselves to low Peclet numbers
where MIPS does not occur.  Our central question is: What are
non-equilibrium signatures in the motion of the passive probe in this
active fluid?

A number of previous studies have asserted that the dynamical behavior
of a probe immersed in an active bath is analogous to that of an
isolated active particle \cite{wu,Volpe,Callegari,steff}. Therefore,
we will also characterize the behavior of an isolated ALP, as a
reference and to identify its characteristic non-equilibrium
signatures. 

The rest of the paper is organized as follows: In the next section, we
introduce the model and give some simulation details. Then, in Section
\ref{sec:single_ALPs}, we will discuss isolated ALPs, for which many
quantities can be calculated analytically. Section
\ref{sec:full_system} describes our characterization of the full
system of an ALP with immersed probe particle. The main result is that
it is {\em not} possible to extract non-equilibrium signatures from
just the knowledge of the unperturbed trajectory of a single probe,
without additional knowledge on bath properties.  However, in Section
\ref{sec:1FDT} we will then identify a manipulation of the probe which
can be used to identify the non-equilibrium character of the system.
We summarize and conclude in Section \ref{sec:conclusions}.

\section{Model and Simulation Details}
\label{sec:syssim}

\begin{figure}[!ht]

  \centering
  \includegraphics[width=.8\linewidth]{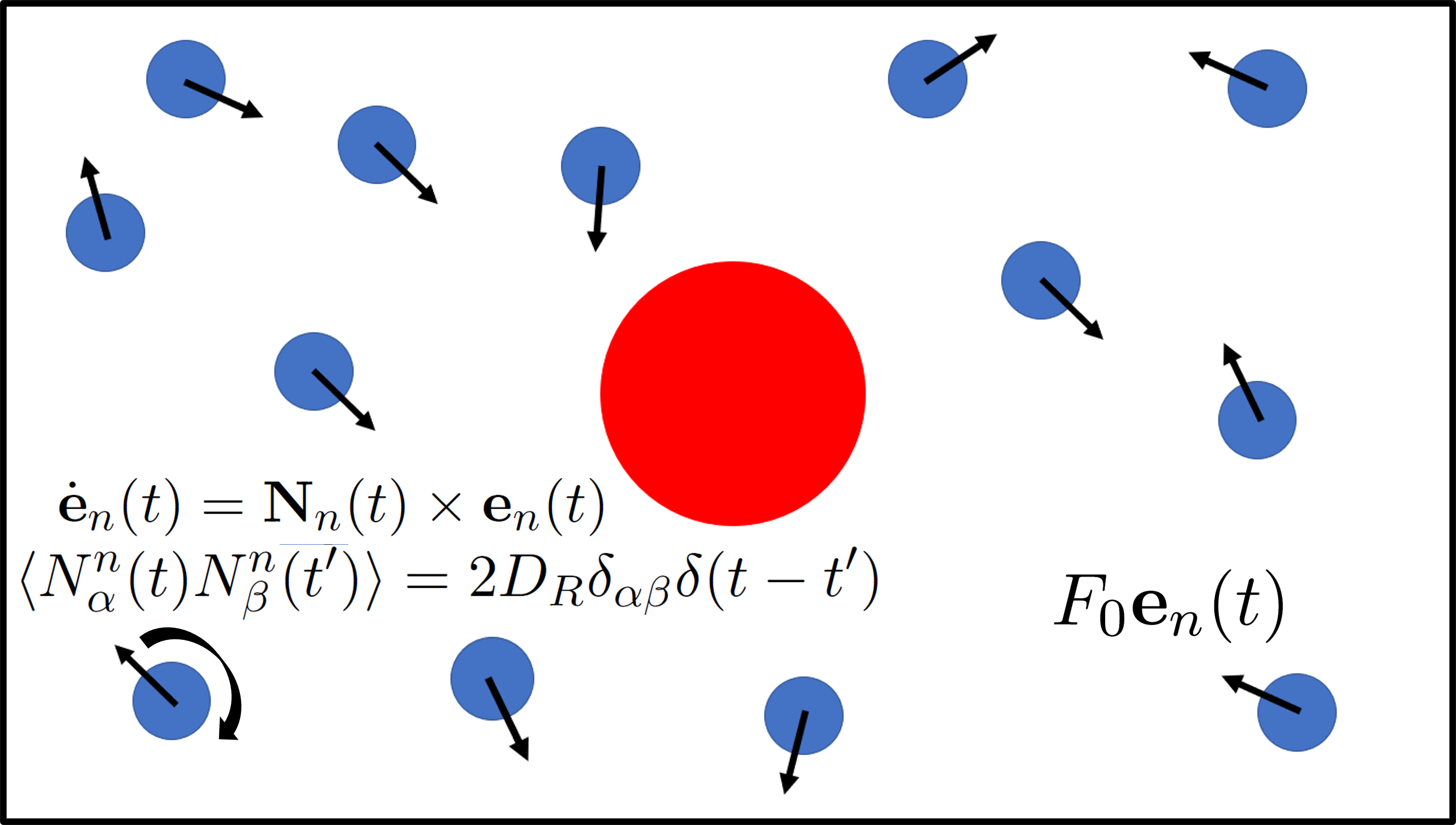}  
\caption{System of a passive probe immersed in a bath of active Langevin particles.}
\label{fig:sys}

\end{figure}

We consider a three dimensional system of a passive probe immersed in
a bath of active Langevin particles (ALPs) of mass $m$ and radius $R$,
which propel themselves with a constant force $F_0$ subject to
rotational diffusion with a diffusion constant
$D_R$~\cite{ABP_ALP,ALPs_Takatori,
Inertial_delay,Hidden_entropy_Marchetti,Enhanced_diff_Marchetti,MITD,Gompper_local_stress,Loewen_TD_inertia}
(see Fig.~\ref{fig:sys}).  The ALPs are coupled to a thermal bath with
temperature $\kbT$ via a Markovian, Langevin thermostat, and they
interact with each other and with the immersed probe by repulsive hard
core interactions of the Weeks-Chandler-Anderson (WCA) type
\cite{WCA}.  

Specifically, we follow Refs.~\cite{ALPs_Takatori,
Hidden_entropy_Marchetti} and set the rotational inertia of ALPs to
zero for simplicity.  The equations of motion for an ALP, $n$, in the
bath, are thus given by
\begin{equation}
\label{eq:alp_int}
\begin{split}
m \dot{\mathbf{v}}_n(t)&=F_0\mathbf{e}_n(t)-\gamma \mathbf{v_n}(t)+\boldsymbol{\xi}_n(t)\\
&-\nabla U_\mathrm{WCA}(\mathbf{r}_n-\mathbf{R})-\sum_{n\neq m} \nabla U_\mathrm{WCA}(\mathbf{r}_n-\mathbf{r}_m),
\end{split}
\end{equation}
where $F_0$ is the propulsion force of the active particle,
$\mathbf{e}(t)$ is the orientation of the active particle, and
$\gamma=6\pi\eta R$ is the damping constant for an ALP radius $R$ in a
thermal bath with viscosity $\eta$. The term $\boldsymbol{\xi}(t)$
represents a stochastic force that mimics implicit collisions of the
ALPs with thermal bath particles. These collisions are modelled as
Gaussian white noise with mean zero and variance given by
a fluctuation-dissipation relation
\begin{equation}
\label{eq:trans_diff}
\langle \xi_i(t)\xi_j(t') \rangle = 2 \gamma k_B T \delta_{ij} \delta (t-t').
\end{equation}
The resulting translational diffusion coefficient of isolated ALPs
is given by $D_T=k_B T/\gamma$. 
Finally, the terms $-\nabla U_\mathrm{WCA}(\mathbf{r}_n-\mathbf{R})$
and $-\sum_{n\neq m} \nabla U_\mathrm{WCA}(\mathbf{r}_n-\mathbf{r}_m)$
in Eq.\ \ref{eq:alp_int} describe the WCA interactions with the probe
and with all other ALPs, respectively. 

The time evolution of the orientation of the ALP, $\mathbf{e}(t)$, is
governed by rotational diffusion, 
\begin{equation}
\label{eq:rot}
\dot{\mathbf{e}}(t)=\mathbf{N}(t) \times \mathbf{e}(t),
\end{equation}
where $\mathbf{N}(t)$ is again Gaussian white noise with a mean of $0$ 
and a variance (another fluctuation-dissipation relation)
\begin{equation}
\label{eq:rot_diff}
\langle N_{\alpha}(t)N_{\beta}(t') \rangle = 2 D_R \delta_{\alpha \beta} \delta (t-t').
\end{equation}
Here $D_R$ is the rotational diffusion constant, which is given by
$D_R=3D_T/4R^2$ for a particle of radius $R$.

The immersed probe only experiences forces from interactions with the
surrounding ALPs. Unlike the ALPs, it is not coupled to the
thermal bath. Thus, the equation of motion for the probe is:
\begin{equation}
\label{eq:eom_coll}
M \dot{\mathbf{V}}(t)=-\sum_n \nabla U_\mathrm{WCA}(\mathbf{R}-\mathbf{r}_n),
\end{equation}
where $M$ is the mass of the probe, $\mathbf{V}(t)$ is its velocity, and $U_\mathrm{WCA}(\mathbf{R}-\mathbf{r}_n)$ is the WCA potential due to an ALP particle, $n$, at position $\mathbf{r}_n$.

All simulations are performed using LAMMPS~[57]. The length, energy,
and mass scales in the system are defined by the Lennard-Jones (LJ)
diameter $\sigma$, energy $\epsilon$, and mass $m$, respectively,
which defines the LJ time scale $\tlj = \sigma
\sqrt{m/\epsilon}$. We use truncated and shifted LJ potentials with
the energy scale $\epsilon$ which are cut off at
$r_{\mathrm{c}}=2^{\frac{1}{6}}\sigma$, resulting in purely repulsive
WCA interactions as described above. The square simulation box has
periodic boundary conditions in all three dimensions and a side length
based on the desired density of the bath. 
$43.0887\sigma$.  The probe has a mass of $M=100m$ and is defined as a
rigid body with a radius $R_p=3\sigma$. The body of the probe is
constructed so that its surface is smooth, resulting in full slip
boundary conditions for the LJ fluid. The active bath consists of ALPs
with a mass of $m_{\mathrm{ALP}}=1m$ and a radius of $R=0.5\sigma$.
The number of ALPs in the bath is also determined by the desired
density of the bath. The parameters of the thermal bath are chosen
such that $\kbT = 1 \; \epsilon$ and 
$\eta = 1 \; m/(\sigma \tlj)$, resulting in
$\gamma = 3 \pi m/\tlj$, $D_T = (3 \pi)^{-1} \sigma^2/\tlj$,
and $D_R = \pi^{-1} /\tlj$.
The driving force $F_0$ is
varied in the range up to $F_0 \le 60 \: m \sigma^2/t_{lJ}$,
resulting in Peclet numbers $\mathrm{Pe}=F_0/(2D_R\gamma R) \le 20$.

In the following, all quantities will be given in simulation units
$\sigma$, $\epsilon$, and $\tlj$. 

\section{Isolated ALPs}
\label{sec:single_ALPs}

We come to the characterization and the non-equilibrium signatures in
the behavior of isolated ALPs.  Analytical results for certain
behaviors of isolated ABPs~\cite{Volpe,ABP_MSD,Christina_Kurzthaler}
and ALPs with rotational inertia~\cite{ABP_ALP,Soudeh_Lowen} are
already known.  Here, we will expand upon these results for ALPs
without rotational inertia. 

To simplify the notation, we will use dimensionless quantities $\htt =
t \: \gamma/m$, $\hv = \mathbf{v} \: \sqrt{{m}/{\kbT}}$, $\hf = F_0
\frac{1}{\gamma} \sqrt{{m}/{\kbT}}$, and $\hdr = D_R \: m/\gamma$. The
equations of motion of an isolated ALP then read

\begin{equation}
\label{eq:eom}
\frac{\ud}{\ud \htt} \hv(\htt) = \hf \mathbf{e}(\htt) - \hv(\htt) + \hxi(\htt); \quad
\frac{\ud}{\ud \htt} \mathbf{e}(\htt) = \hN(\htt) \times \mathbf{e}(\htt)
\end{equation}
where $\mathbf{e}(\htt)$ is a unit vector and the stochastic contributions $\hxi$ and $\hN$ satisfy fluctuation dissipation relations $\langle \hxi(\htt) \hxi(\htt') \rangle = 2 \: \unity \delta(\htt - \htt')$ and $\langle \hN(\htt) \hN(\htt') = 2 \hdr \unity \delta(\htt - \htt')$. From these equations, we can analytically calculate the velocity autocorrelation function (VACF) $C_{\hv}(\htt) = \langle \hv(\htt')~\cdot~\hv(\htt' + \htt) \rangle$ of isolated ALPs (see Appendix~\ref{sec:supp_calc_vacf}): 
\begin{equation}
\label{eq:vv_alp}
    C_{\hv}(\htt) = 3 \ue^{- |\htt|} + \frac{\hf^2}{1-4 \hdr^2}\left(\ue^{-2 \hdr |\htt|} - 2 \hdr \ue^{-|\htt|} \right).
\end{equation}

\begin{figure}
  \centering
  \includegraphics[width=0.9\linewidth]{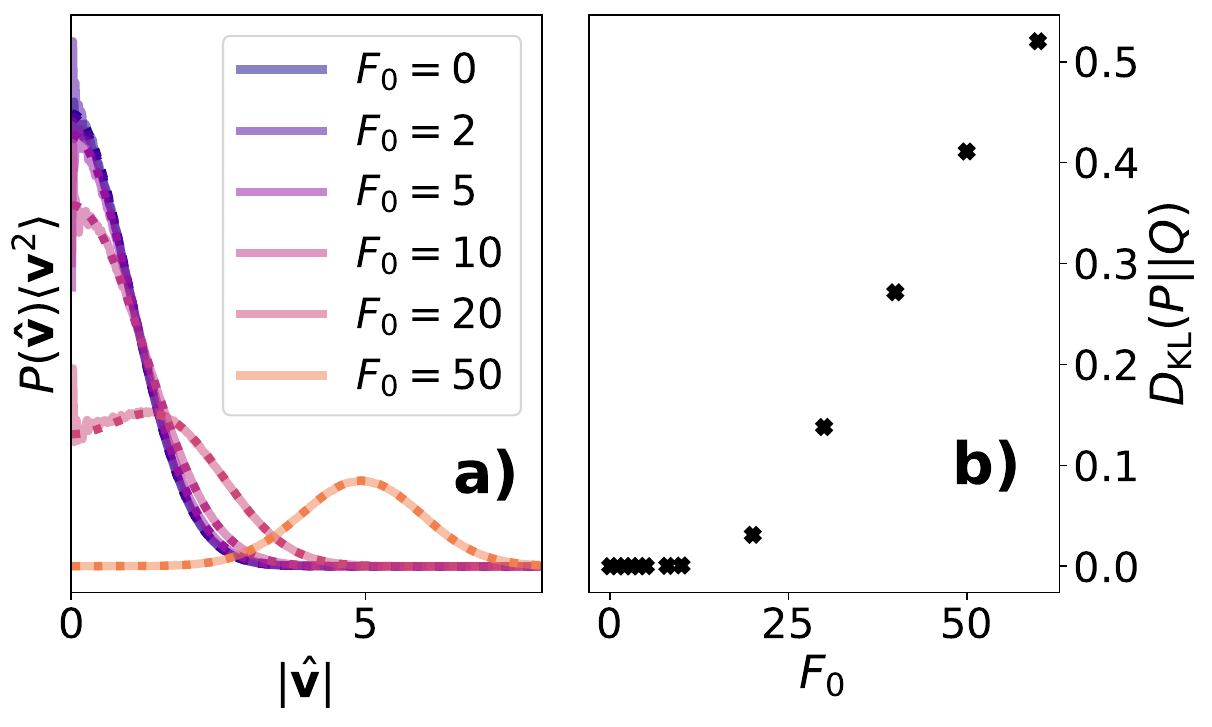}  
\caption{\textbf{a)} Velocity distribution for isolated ALPs for different driving forces $F_0$ as indicated. Solid lines show simulation data, dashed lines the prediction of Eq. (\ref{eq:vd_probability}).
The curves have been rescaled by $\langle \mathbf{v}^2 \rangle$ for better visibility. \textbf{b)} Relative entropy (Eq.~\eqref{eq:re}) between the distributions in a) and a Gaussian distribution 
with same standard deviation as a function of $F_0$. 
}
\label{fig:vd_re_iso}
\end{figure}

In the case of a passive particle in equilibrium, the initial value of the VACF is related to its temperature through the equation $\langle\mathbf{v}^2\rangle=dk_\mathrm{B}T/m$, where $d$ is the number of spatial dimensions in the system~\cite{Zwanzig_Non}. In analogy to this relation, we define the kinetic temperature of an ALP as $k_\mathrm{B}T_\mathrm{eff}=m\langle\mathbf{v}^2\rangle/d$. Unlike a passive particle, however, $T_\mathrm{eff}$ of an ALP will not necessarily be equal to the temperature of the thermal bath, $T$. In fact, from Eq.~\eqref{eq:vv_alp}, we find that it is given by:
\begin{equation}
\label{eq:v2_alp}
\frac{\kbT_\mathrm{eff}}{\kbT} = 1 + \frac{1}{3} \: \frac{\hf^2}{(1+2 \hdr)}.
\end{equation}
Therefore, we expect $(\kbT_\mathrm{eff}-\kbT)\propto F_0^2$. This scaling relation is analogous to that between the effective diffusion constant of an ABP and its active velocity~\cite{Volpe}.

\begin{figure*}[!ht]
  \centering
  \includegraphics[width=.85\linewidth]{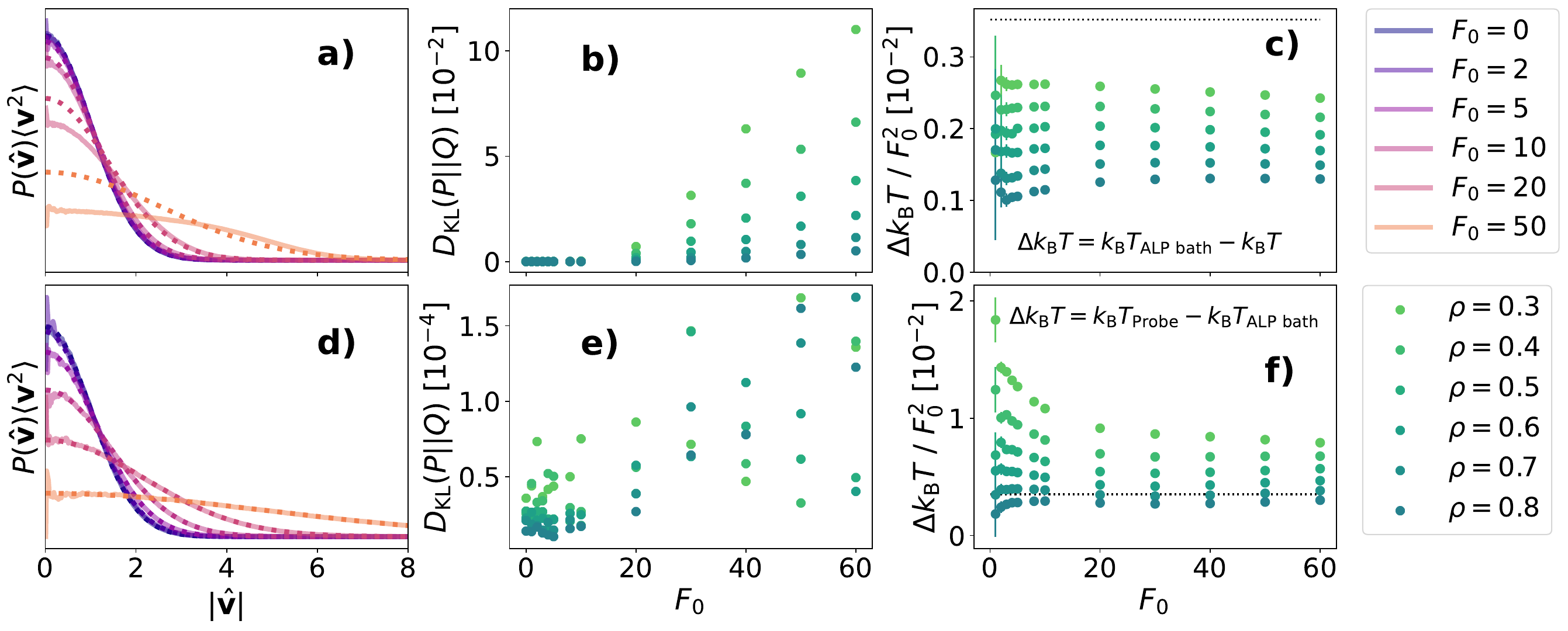}  
\caption{\textbf{a,d)}: Velocity distribution for \textbf{a)} bath ALPs and \textbf{b)} the immersed probe from simulation data (solid lines) at $\rho=0.4$ for different driving forces $F_0$, as indicated. The dotted lines show Gaussian distributions with the same standard deviation for comparison. \textbf{b,e)}: Corresponding relative entropies (Eq.~(\ref{eq:re})) between the simulation data and Gaussians with the same standard deviation. \textbf{c,f}): Kinetic temperature difference scaled by $\hf^2$ between $\textbf{c)}$ a bath ALP and the thermal bath, and $\textbf{f)}$ the probe and the ALP fluid. The dotted black line shows the difference between the kinetic temperature of an isolated ALP and its thermal bath.
\label{fig:vd_re_temp}}
\end{figure*}

The failure of an ALP to equilibrate to the same temperature as its thermal bath is a hallmark non-equilibrium signature; however, knowledge of the ALP thermostat is required to observe this enhanced temperature. From simulation data, we additionally observe that as $F_0$ of an ALP increases, its velocity distribution becomes increasingly non-Gaussian, as shown in Fig.~\ref{fig:vd_re_iso}a) for velocity distribution $P(\hv)$, a description of which is in Appendix~\ref{sec:supp_calc_pv}.
This can also be inferred from the Fokker-Planck equation corresponding to Eq.\ (\ref{eq:eom}), which 
predicts a stationary velocity distribution 
\begin{equation}
    P(\hv) = {\cal N} \: \ue^{- \hv^2/2} \: 
    \exp(\hf \: U(|\hv|))
\end{equation}
in the limit $\hdr \to 0$ (see Appendix~\ref{sec:supp_calc_pv}), 
where $U'(|\hv|) = \langle \cos \sphericalangle(\hv,\mathbf{e})
\rangle_{\hv}$ is the mean cosine of the angle between $\hv$ and
$\mathbf{e}$ for given ALP velocity $\hv$. The first factor
$\ue^{-\hv^2/2}$ represents  a Maxwell distribution at temperature
$k_B T$, the second the effect of the propelling force. In the limit
$\hdr \to 0$, the ALPs rotate infinitely slowly, and $U(|\hv|)$ can be
evaluated as $U(|\hv|)= \frac{1}{\hf} \ln\big(\sinh(\hf \hv)/\hf
\hv\big)$ (see Appendix~\ref{sec:supp_calc_pv}).
Motivated by this result, we conjecture that $P(\hv)$ can, in general,
be approximated by the expression,
\begin{equation}
\label{eq:vd_probability}
    P(\hv) \approx {\cal N} \: \ue^{- \hv^2/2} \: 
    \frac{\sqrt{1+2 \hdr}}{\hf |\hv|} \sinh \Big(\frac{\hf |\hv|}{\sqrt{1+2 \hdr}}\Big),
\end{equation}
which reproduces the correct limits $\hdr \to 0$ and $\hf \to 0$ and the correct kinetic temperature, Eq.\ (\ref{eq:v2_alp}).  Fig.\ \ref{fig:vd_re_iso}a) shows that the simulation data (solid lines) are very well captured by this Ansatz (dashed line). 

The non-Gaussianity of the velocity distribution is another signature
of the non-equilibrium nature of an isolated ALP, and one which can be
extracted from the ALP trajectory alone. To better quantify this
deviation from a Gaussian distribution, we calculate the relative
entropy (Kullback-Leibler divergence) between the velocity
distribution of an isolated ALP, as calculated from simulation data,
and a zero-centered Gaussian distribution whose standard deviation
corresponds to the theoretical temperature given in
Eq.~\eqref{eq:v2_alp}. The relative entropy between these two
distributions is defined as:
\begin{equation}
\label{eq:re}
D_\mathrm{KL}(\mathcal{P}(\hv)||\mathcal{Q}(\hv))= \int_{\infty}\!\! \ud^3 \hat{v} \: \mathcal{P}(\hv)\ln{\Bigg(\frac{\mathcal{P}(\hv)}{\mathcal{Q}(\hv)}\Bigg)},
\end{equation}
where $\mathcal{P}(\hv)$ is the velocity distribution from simulation
data and $\mathcal{Q}(\hv)$ is the reference Gaussian distribution.
This relative entropy is graphed for an isolated ALP as a function of
$F_0$ in Fig.~\ref{fig:vd_re_iso}b). We find that the velocity
distribution of an ALP remains approximately Gaussian until an active
force of $\sim10$, at which point it becomes increasingly
non-Gaussian. 

\section{ALP Fluids and Immersed Probe}
\label{sec:full_system}

\subsection{Velocity Distributions and Autocorrelation Functions}

With this knowledge of the behavior of an isolated ALP, we now
investigate the full system of a passive probe particle immersed in an
ALP fluid with varying ALP number density $\rho$ -- focusing on the
search for non-equilibrium signatures in the behavior of the immersed
probe, and that of a randomly selected interacting ALP from the bath.
The immersed probe particle has the mass $M=100 m$ and the radius
$R_p=6 R$. Unless stated otherwise, it only experiences forces from
interactions with the surrounding ALPs and is not coupled to the
thermal bath.

Similar to isolated ALPs, bath ALPs exhibit the non-equilibrium
signature of a non-Gaussian velocity distribution for $F_0\gtrsim10$.
This is apparent from Fig.~\ref{fig:vd_re_temp}a),  which compares
their velocity distribution from simulation data to a Gaussian
distribution with the same standard deviation, and in
Fig.~\ref{fig:vd_re_temp}b), which shows the relative entropy between
these two curves. The velocity distribution of the immersed passive
particle, on the other hand, remains Gaussian at all values of $F_0$,
independent of both the activity of the bath ALPs and their density
(see Fig.~\ref{fig:vd_re_temp}d) and e)). Thus, in contrast to the
case for both isolated and bath ALPs, the velocity distribution of the
probe particle does not reveal the non-equilibrium nature of the
system.

The immersed probe does, however, still exhibit the non-equilibrium
signature of an enhanced kinetic temperature. In fact, for an ALP bath
with a given $F_0$, we find that the kinetic temperature of the
immersed probe is even {\em higher} than that of the ALP bath
(Fig.~\ref{fig:vd_re_temp}f)), which is already enhanced from that of
an equilibrium bath (Fig.~\ref{fig:vd_re_temp}c)). Thus, this system
has a twofold lack of thermalization: firstly, 
between the ALP bath and the thermal bath, and secondly, 
between the probe and the ALP bath. Remarkably, we find these effects
persist even when the probe is thermostatted
(Fig.~\ref{fig:temp_thermo} of the Appendix). The fact that the probe
does not equilibrate to the same kinetic temperature as the ALP bath
clearly reveals the non-equilibrium nature of the system; however, it
requires analysis of both bath and probe. To uncover non-equilibrium
signatures by solely observing the probe, we further investigate its
dynamic properties.

\begin{figure}
  \centering
  \includegraphics[width=1\linewidth]{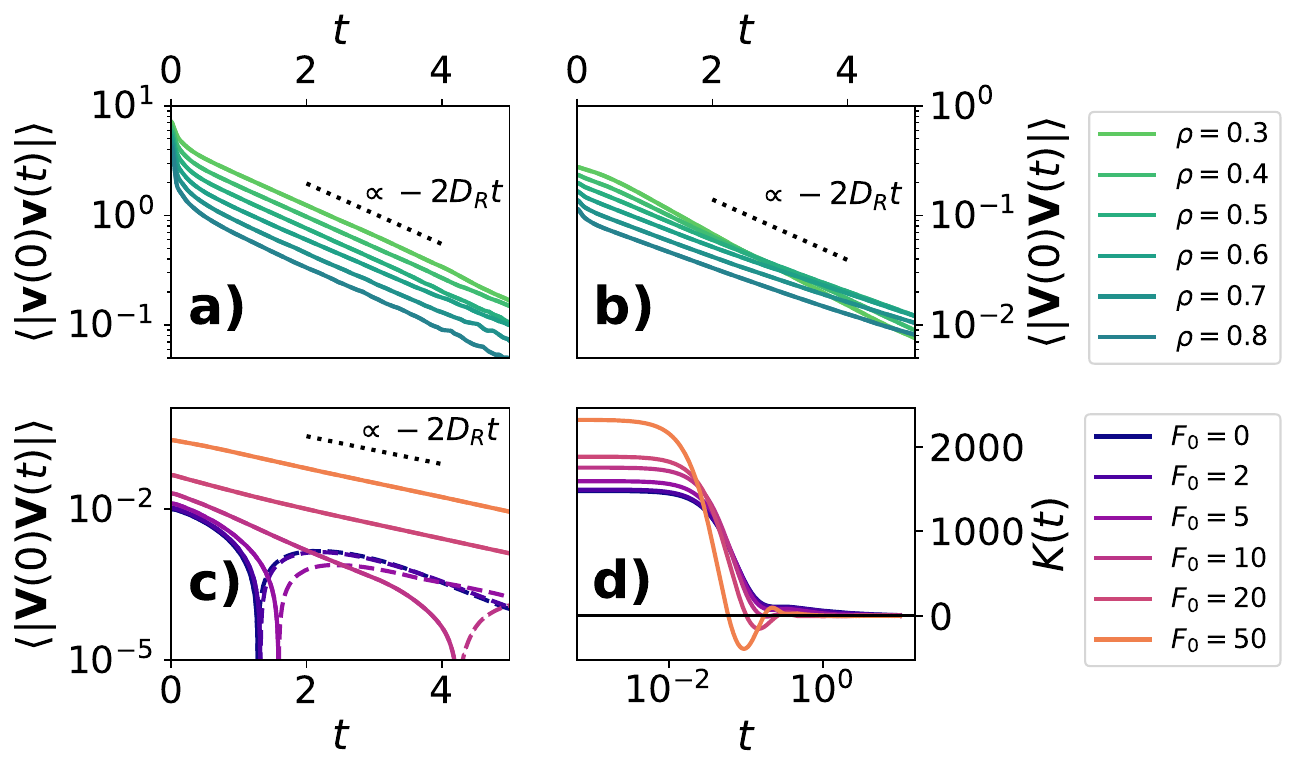}  
\caption{VACF and memory kernel. \textbf{a)}, \textbf{b)} Semi-logarithmic plot of the VACF for different densities with $F_0=50$ of \textbf{a)} a bath ALP, \textbf{b)} the immersed probe. \textbf{c)} Semi-logarithmic plot of the VACF for the immersed probe in a bath of density $\rho=0.4$ for different values of $F_0$. Dashed lines show the absolute value of the VACF. \textbf{d)} Semi-logarithmic plot of the memory kernel of the immersed passive particle in a bath of $\rho=0.4$ for different values of $F_0$.}
\label{fig:vvs}
\end{figure}

We first focus on the VACF, which we determined analytically for an
isolated ALP in Eq.~\eqref{eq:vv_alp}. As is the case for the VACF of
an isolated ALP (assuming $\gamma/m>2D_R$), in the limit $t\to\infty$,
the VACF of a bath ALP decays exponentially as $2D_R$, independent of
density, as is shown in Fig.~\ref{fig:vvs}a). However, both for the
bath ALPs and the probe, this exponential decay is only observed at
sufficiently high values of $F_0$, as is shown for the probe in
Fig.~\ref{fig:vvs}c).

Despite the significant size difference between an ALP and the probe,
the VACF of the probe decays exponentially at a very similar rate to
that of an isolated ALP, as shown in Fig.~\ref{fig:vvs}b). Thus, the
probe acquires some properties of the ALPs in the surrounding bath.
However, unlike the VACF of  bath ALPs, the VACF of the probe in the
long-time limit depends on the density of bath ALPs.

\subsection{Non-Markovian Behavior of the Probe}

Although the behavior of the VACF of a probe in an active bath differs
from that of a probe in a passive bath, these differences are not
inherent to non-equilibrium. Thus, to continue searching the dynamic
behavior of the probe for non-equilibrium signatures, we map its
movement onto the generalized Langevin equation
(GLE)~\cite{Zwanzig_Non}:
\begin{equation}
\label{eq:gle}
M \dot{\mathbf{V}}(t)=-\int^t_0 \mathrm{d}s \: K(t-s) \mathbf{V}(s)+\bm{\Gamma}(t),
\end{equation}
where $M$ is the mass of the probe, $\mathbf{V}(t)$ is its velocity,
$K(t-s)$ is its memory kernel, and $\bm{\Gamma}(t)$ is the stochastic
force on the probe. In mapping the motion of the probe to the GLE, we
explicitly allow that the effective dynamics of the colloid in the ALP
fluid may be non-Markovian. The memory kernel $K(t)$ is determined
from the VACF by Volterra inversion, as suggested by the Mori-Zwanzig
projection operator
formalism~\cite{Zwanzig_Non,MZ,Zwanzig_Mem,Shin_2010,memory_review}.
We emphasize that the GLE is a coarse-grained model equation; the true
dynamical equation of motion for the probe in the explicit active
fluid is given in Eq.~(\ref{eq:eom_coll}).

As shown in Fig.\ \ref{fig:vvs}d), the shape of the memory kernel
changes qualitatively with increasing activity level in the ALP fluid.
It becomes non-monotonous and negative at intermediate times,
indicating transient positive feedback that promotes superdiffusive
behavior. This is also observed in the memory kernel of isolated ALPs
as obtained from Eq.~(\ref{eq:vv_alp}) (see
Appendix~\ref{sec:supp_calc_vacf}), 
\begin{equation}
K_{_{\textrm{ALP}}}(t) = \frac{1}{C_{\hv}(0)}
\big(6 \: \delta(t) - \beta \ue^{-t/\tau}\big) 
\end{equation}
with 
\begin{equation}
\tau = \frac{C_{\hv}(0)}{(\hf^2 + 6 \hdr)} \quad \mbox{and}  \quad
\beta = \frac{(\tau -1)}{\tau (1+2 \hdr)} \: \hf^2. 
\end{equation}
We conclude that the dynamics of the probe particle does acquire some
properties of ALPs, as claimed in Refs.\
\cite{wu,Volpe,Callegari,steff}. However, memory kernels may also
oscillate in equilibrium systems
\cite{Carof_2014,Jung_2017,Bockius_21}, hence this cannot be used to
diagnose non-equilibrium. 

At equilibrium, the dissipative interactions, described by the memory
kernel, can be related to the stochastic forces through the relation
$\langle \bm{\Gamma}(0) \bm{\Gamma}(t) \rangle = m \langle
|\mathbf{v}|^2\rangle K(t)$~\cite{Zwanzig_Non}. This relation is one
type of fluctuation dissipation relation, which we will henceforth
refer to as the $2^{nd}$FDT~\cite{wip}.  Several studies of a passive
probe immersed in an active bath have concluded that the
non-equilibrium nature of the bath leads to a violation of the
$2^{nd}$FDT for the immersed
probe~\cite{Chen,Lau,steff,maggi_fdt,Argun,maggi}, which has been
touted as a signature of the non-equilibrium nature of this system. In
seeming contradiction, Loi et al.~\cite{Loi} found that the
$2^{nd}$FDT for the correlation function of the independent scattering
function is fulfilled for a passive probe in an active bath. This
apparent contradiction is due to a difference in definition of the
stochastic and dissipative forces on the probe. In fact,
Ref.~\cite{wip} has shown that whenever the stochastic and dissipative
forces on the probe are defined according to the Mori-Zwanzig
projection operators~\cite{Zwanzig_Non,MZ,Zwanzig_Mem,proj_op}, then
the second fluctuation dissipation theorem will be fulfilled exactly,
even if the system is out of equilibrium. Therefore, violation of the
$2^{nd}$FDT is not an inherent trait of the non-equilibrium nature of
the bath, but rather a matter of definition.  Indeed, in accordance
with the results of Loi et al.~\cite{Loi} and Ref.~\cite{wip},  we
find that the $2^{nd}$FDT for the GLE is fulfilled exactly for the
immersed probe for all driving forces $F_0$ (Fig.~\ref{fig:fdt_conf}). 



Among other non-equilibrium signatures proposed in the literature~\cite{Cerasoli_Spectral_Fingerprints,Gonzalez_Exp_DBViolation,Seifert_Dissipation_Colloidal,Roldan_Est_Diss,Manikandan_2021}, one of the most ubiquitous is entropy production associated with a difference of  probabilities between forward and time-reversed trajectories, which has been found in a number of active systems \cite{Sunghan_Stefano,Hidden_entropy_Marchetti,Dabelow_Irr_ActMatt}. However, in our probe trajectories, we found no evidence of such a breaking of time reversal symmetry (see Appendix~\ref{sec:entropy}).

\section{Response to Small Perturbations}
\label{sec:1FDT}

In equilibrium systems, the response of quantities to infinitesimal
external perturbations can be described by linear response theory.  As
a last test of non-equilibrium signatures in our active system, we
investigate whether the predictions of linear response theory still
hold here. Specifically, we investigate what we will refer to as the
first fluctuation dissipation theorem ($1^{st}$FDT)~\cite{wip}.  It is
derived from linear response theory and relates the non-equilibrium
response of an observable to an infinitesimal perturbation to
the relaxation of fluctuations in the equilibrium system. 
This is in contrast to the $2^{nd}$FDT, which connects
the dissipative interactions in a system with the stochastic noise in
that same system. For a classical system, the $1^{st}$FDT is given
as~\cite{Zwanzig_Non} 

$\chi(t)=-\frac{1}{k_\mathrm{B}T}\frac{\mathrm{d}}{\mathrm{d}t}C_{XY}(t)\Theta(t)$,
where $\Theta(t)$ is the Heaviside function and $C_{XY}(t)$ is the equilibrium correlation function between observables $X$ and $Y$. It has been previously shown that the $1^{st}$FDT will hold even in a non-equilibrium steady-state, provided certain assumptions which imply that the system remains close to equilibrium \cite{1FDT_1972,1FDT_2009,1FDT_2010}.

To test the validity of the $1^{st}$FDT for the immersed passive
particle, we apply a perturbation by kicking it slightly at time $t=0$, i.e., we add an instantaneous force of the form $\alpha(t)=MV_0\delta(t)$, 
where $M$ is the immersed particle mass and $V_0$ is small. 
We then investigate the response of the velocity~\cite{Ciccotti_1FDT,wip}. 
In an equilibrium system, it would be given by
\begin{equation}
\label{eq:1fdt_spec}
\frac{\delta\langle V(t)\rangle}{V_0}=\frac{C^{SS}_V(t)}{C^{SS}_V(0)},
\end{equation}
where $C^{SS}_V(t)$ is the steady state (equilibrium) VACF. The
question is whether this relation still holds if the system is
perturbed from a non-equilibrium steady state such as our active
system.

To test the relation for the immersed passive particle, we perform
two simulations simultaneously: one in which the perturbation 
$\alpha(t)=MV_0\delta(t)$ is applied to the immersed particle (pert)
and one in which no perturbation is applied to the immersed particle
(unpert). Aside from this perturbation, the simulations are identical;
in particular, the same random numbers are used in the thermostat. We
then calculate the response as as $\delta
V(t)=V_{\mathrm{pert}}(t)-V_{\mathrm{unpert}}(t)$ and average this
function over many systems to reduce statistical noise. This method is
identical to that used in Ref.~\cite{Bockius_21,1FDT_2010}. 
To ensure that we are within the regime where the response is linear
in $V_0$, we test multiple values of $V_0$ ($V_0=0.2,~0.4,~0.5$) and
verify that the response function is the same for all values.

\begin{figure}
  \centering
\includegraphics[width=1\linewidth]{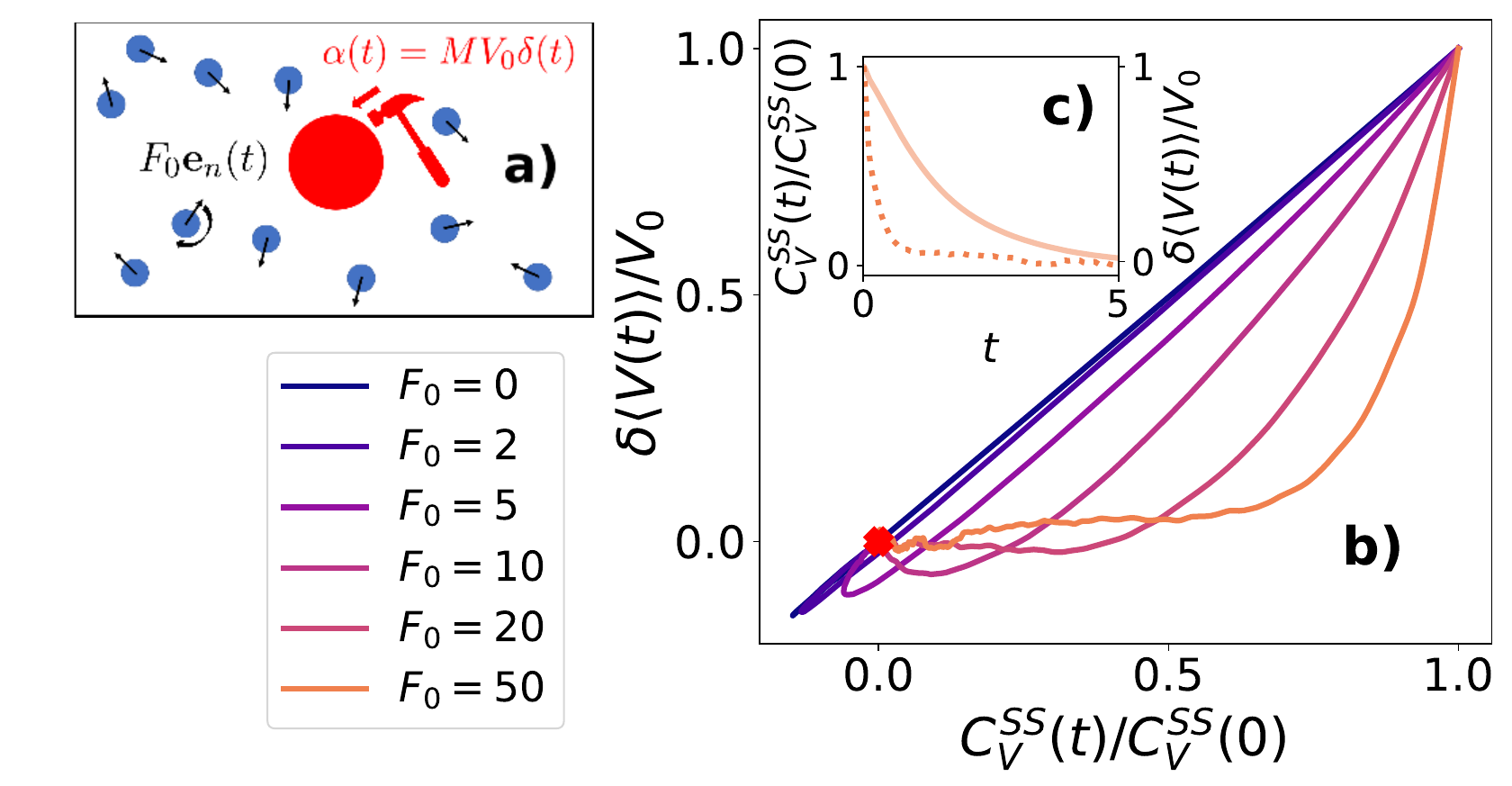}
\caption{Test of the $1^{st}$FDT for the immersed passive particle in a bath with density $\rho=0.4$ for perturbation force $\alpha(t)=MV_0\delta(t)$ with $V_0=0.5$. \textbf{a)} Schematic diagram of the perturbation applied to the probe. \textbf{b)} Response function of the probe velocity versus normalized VACF. The red diamond shows the point (0,0) to which both the functions decay in the long-time limit. If the $1^{st}$ FDT is fulfilled, the curve is a straight line along the diagonal. \textbf{c)} Response function of the probe velocity (dotted line) and normalized VACF (solid line) as a function of time for $F_0=50$.}
\label{fig:fdt1}
\end{figure}

The results for $\delta \langle V(t) \rangle/V_0$ are summarized in
Fig.~ \ref{fig:fdt1} and plotted versus $C_V^{SS}(t)/C_V^{SS}(0)$.  If
Eq.\ (\ref{eq:1fdt_spec}) holds, the data in such a plot should 
all collapse onto one diagonal line. Fig.~\ref{fig:fdt1} shows that this is
indeed the case at driving force $F_0 = 0$ (an equilibrium bath),
but deviations are observed already at low values of $F_0$.  As the
value of $F_0$ increases, thereby driving the system further out of
equilibrium, the violation of the $1^{st}$FDT becomes increasingly
large. Thus, even though the immersed passive particle reaches a
nonequilibrium steady state, the $1^{st}$FDT is {\em not} fulfilled
for a passive particle immersed in an active bath. This effect can
be used to detect non-equilibrium even if one has only access to
the observation of a single probe particle.

\section{Conclusions}

\label{sec:conclusions}

To summarize, we have shown that non-equilibrium signatures of a
passive particle in an active bath are not as readily available as
those of an active particle itself. Although the immersed probe does
inherit some properties of the active bath, such as slowly decaying
velocity correlations and a partially negative memory kernel, unlike
an active bath particle, it does not inherit a non-Gaussian linear
velocity distribution. The probe acquires an enhanced kinetic
temperature, as does the active bath, but these two temperatures are
not equal: the temperature of the immersed probe is even higher than
that of the active bath. Therefore, the lack of thermalization in this
system is twofold.
The failure of the probe to equilibrate to the same temperature as the
active bath is one non-equilibrium signature of the system; however,
detecting this enhanced temperature requires knowledge of the bath
itself. As one way to detect a non-equilibrium signature from only the
trajectory of the probe, we propose to slightly perturb the probe. If
the bath is not at equilibrium, this will then be revealed through a
violation of the first fluctuation dissipation theorem.


\section*{Conflicts of interest}
There are no conflicts to declare.

\section*{Appendix}
\appendix

\section{Properties of isolated ALPs}
\label{sec:supp_calc_vacf}
In rescaled quantities $\hf$ and $\hdr$, 
the equation for the velocity of a non-interacting ALP is:
\begin{equation}
\label{eq:supp_v_alp}
\hv(\htt) = \hv(0) \ue^{- \htt} + \hf \int_0^{\htt} \ud s\: \ue^{-(\htt - s)}
\mathbf{e}(s) + \int_0^{\htt} \ud s \:\ue^{- (\htt - s)} \hxi(s)
\end{equation}
We now use this equation to derive the theoretical VACF of an ALP. We know that the VACF can be calculated as:
\begin{equation}
\label{eq:supp_vacf}
\langle\hv(\htt)\hv(\htt')\rangle=\frac{1}{T}\int_0^T\mathrm{d}s \: \hv(\htt+s) \: \hv(\htt'+s),
\end{equation}
in the limit $T\to\infty$. We now assume that the initial time is in the infinite past, as done in~[31], so that the instantaneous velocity is only determined by the noise terms in Eq.~\ref{eq:supp_v_alp}. Under this assumption, our ALP instantaneous velocity is:
\begin{equation}
\label{eq:supp_v_alp_inf}
\hv(\htt)=\hf\int^\infty_0 \!\!\! \ud u \: \ue^{-u}\: \mathbf{e}(\htt-u)
+ \int_0^\infty\ud u' \ue^{-u'} \: \hxi(\htt-u'),
\end{equation}
where we have slightly rearranged the time integral for convenience. Following Eq.~\eqref{eq:supp_vacf}, we then multiply this velocity by the velocity at some other time ($\hv(\htt')$), average over the noise, and integrate over the time interval $T$ in the limit $T\to\infty$. From Eqs.~\eqref{eq:trans_diff},~\eqref{eq:rot}, and \eqref{eq:rot_diff} we know that $\langle\mathbf{e}(\htt)\hxi(\htt')\rangle=0$; therefore, the second and third terms are zero. We also know that $\langle\mathbf{e}(\htt)\mathbf{e}(\htt')\rangle=\exp{(-2\hdr|\htt-\htt'|)}$ and $\langle\hxi(\htt)\hxi(\htt')\rangle=2\kbT \delta{(\htt-\htt')}$. Therefore, performing all the above integrations, our equation for the VACF of an isolated ALP is:
\begin{equation}
\label{eq:supp_vv_alp}
\begin{split} 
\langle \hv(\htt) \hv(\htt') \rangle = &
   3 \ue^{- |\htt-\htt'|} \\
  & + \frac{\hf^2}{1-4 \hdr^2}\left(\ue^{-2 \hdr |\htt-\htt'|} - 2 \hdr \ue^{-|\htt-\htt'|} \right),
\end{split}
\end{equation}
which is given as Eq.~(\ref{eq:vv_alp}) in the main text. The corresponding memory kernel $K(t)$ which solves the Volterra equation (see below)
\begin{equation}
   \label{eq:supp_volterra_alp}
    \frac{\ud}{\ud \htt} \langle \hv(0) \hv(\htt) \rangle
    = - \int_0^{\htt} \ud s \: K(t-s) \langle \hv(0) \hv(s) \rangle
\end{equation}
can be calculated by Fourier methods, giving
\begin{equation}
\label{eq:supp_memory_alp}
    K_{_{\text{ALP}}}(\htt) = \frac{6}{\langle \hv^2 \rangle} \Big(
    \delta(\htt) - \frac{\tau-1}{6 \tau} \: \frac{\hf^2}{1 + 2 \hdr} \: \ue^{-t/\tau} \Big),
\end{equation}
with $\tau = \langle \hv^2 \rangle/(\hf^2 + 6 \hdr)$ as quoted in the main text. To verify this result, one can simply insert \eqref{eq:supp_memory_alp} and \eqref{eq:supp_vv_alp} into Eq.~\eqref{eq:supp_volterra_alp}.

From Eq.~\eqref{eq:supp_vv_alp}, we can then find the mean squared velocity of an ALP by taking $t=t'$:
\begin{equation}
\label{eq:supp_v2_final}
\langle\hv^2\rangle= 3 +\frac{\hf^2}{1+2 \hdr}.
\end{equation}
The temperature of a passive particle in equilibrium is directly proportional to the mean-squared velocity through the equation $\langle\mathbf{v}^2(t)\rangle=dk_\mathrm{B}T/m$, where $d$ is the number of dimensions in the system~[31]. From this equation, we define the kinetic temperature of an ALP as $k_\mathrm{B}T=m\langle\mathbf{v}^2(t)\rangle/d$. Therefore, we can use Eq.~\eqref{eq:supp_v2_final} to calculate the kinetic temperature of an ALP, which is listed as Eq.~(\ref{eq:v2_alp}) of the main text in terms of rescaled quantities $\hf$ and $\hdr$.

The stochastic equations of motion for an isolated ALP, Eq.\ (\ref{eq:eom}) in the main text, are equivalent to a Fokker-Planck equation for the time-dependent distribution function $\textrm{P}(\hv,\mathbf{e},\htt)$:
\begin{equation}
    \label{eq:supp_fp_alp}
    \begin{split}
    \frac{\partial}{\partial \htt} \textrm{P}(\hv, \mathbf{e}, \htt)
    = &- \nabla_{\hv} \:  (\hv + \nabla_{\hv} - \hf \mathbf{e}) \: 
         \textrm{P}(\hv,\mathbf{e},\htt)
       \\ & - \hdr \: \Delta_{\mathbf{e}} \: \textrm{P}(\hv,\mathbf{e},\htt),
    \end{split}
\end{equation}
where $\Delta_{\mathbf{e}}$ denotes the angular part of the Laplace operator.
The stationary solution of this equation, $\textrm{P}_s(\hv,\mathbf{e})$, solves
 \begin{equation}  
  \label{eq:supp_fpstationary_alp}
   0  = \Big[\nabla_{\hv} \:  (\hv + \nabla_{\hv} - \hf \mathbf{e}) \: 
          + \hdr \: \Delta_{\mathbf{e}} \Big]\: \textrm{P}_S(\hv,\mathbf{e}).
\end{equation}
In the limit $\hdr \to 0$, this equation is solved by
\begin{equation}
    \textrm{P}_s(\hv,\mathbf{e}) = {\cal N} \: 
    \exp\big( - \frac{\hv^2}{2} + \hf \: (\mathbf{e} \cdot \hv) \big),
\end{equation}
where ${\cal N}=\frac{1}{2 \pi^{3/2}\sqrt{2}}\ue^{-\hf^2/2}$ is the normalization constant. Taking the average over all orientations $\mathbf{e}$ gives the velocity distribution at $\hdr\to0$
\begin{equation}
\label{eq:supp_pv}
    P(\hv) = \int \ud^2 \mathbf{e} \: \textrm{P}_S(\hv,\mathbf{e})
    \propto \ue^{-\hv^2/2} \: \frac{1}{\hf |\hv|} \: \sinh(\hf |\hv|),
\end{equation}
which is quoted above Eq.~(4) of the main text above. The mean squared velocity calculated from this distribution is given by $\langle \hv^2 \rangle = 3 + \hf^2$, which has the same functional form than \eqref{eq:supp_v2_final} with $\hf/\sqrt{1+2\hdr}$ replaced by $\hf$. Based on this observation, we conjecture that $P(\hv)$ at finite $\hdr$ has the same functional form than \eqref{eq:supp_pv}  with $\hf$ replaced by $\hf/\sqrt{1 + 2 \hdr}$. This results in Eq.\ (\ref{eq:vd_probability}).

\section{Calculation of $P(\hv)$}
\label{sec:supp_calc_pv}
To calculate $P(\hv)$ from our simulation data, we calculate the absolute velocity, $|\mathbf{v}|$, of the particle for each time step. We then assign this value of $|\mathbf{v}|$ to an appropriate bin, each of length $dv$, to find the absolute velocity distribution $N[|\mathbf{v}|]$. We divide each of these bins by its true volume, $\delta V=\frac{4}{3} \pi\Big((v+\frac{dv}{2})^3-(v-\frac{dv}{2})^3\Big)$, and scale the distribution by a factor of $\sqrt{{m}/{\kbT}}$ to find $P(\hv)$. The distribution calculated from simulation data is normalized such that $\sum_{|\hv|} P(\hv)\delta V=1$. Theoretical distributions of $P(\hv)$ are normalized such that $\int_\infty\mathrm{d}\hat{v}~4\pi \hat{v}^2 P(\hv)=1$. 

\section{Effective temperature of thermostatted probe}
\label{sec:temp_thermo}

We calculate the kinetic temperature of a thermostatted probe in an active bath in order to determine if the higher temperature of the probe, as compared with an isolated ALP, is due to the lack of thermostatt on the probe. The thermostatted probe is described by the equation of motion:

\begin{equation}
\label{eq:eom_coll_thermo}
M \dot{\mathbf{V}}(t)=-\gamma_P \mathbf{V}(t)+\boldsymbol{\xi}(t)-\sum_n \nabla U_\mathrm{WCA}(\mathbf{R}-\mathbf{r}_n),
\end{equation}
where $M$ is the mass of the probe, $\mathbf{V}(t)$ is its velocity, $\gamma_P=6\pi\eta R_P$ is the damping constant for a probe radius $R_P$ in a thermal bath with viscosity $\eta$, and $U_\mathrm{WCA}(\mathbf{R}-\mathbf{r}_n)$ is the WCA potential due to an ALP particle, $n$, at position $\mathbf{r}_n$. The stochastic noise, $\boldsymbol{\xi}(t)$, is described as in Eq.~\eqref{eq:trans_diff}, except that $\gamma$ is replaced by $\gamma_P$.

\begin{figure}[h]
  \centering
  \includegraphics[width=.8\linewidth]{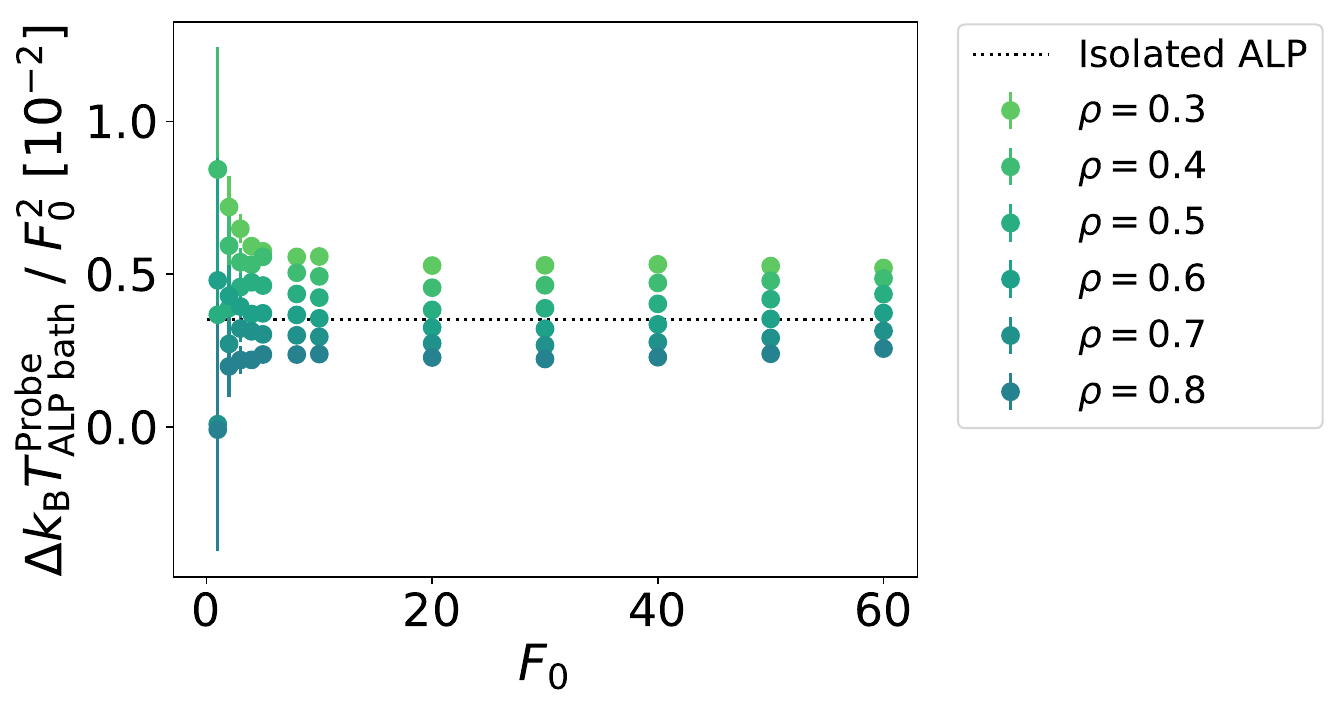}  
\caption{Difference between the kinetic temperature of a thermostatted immersed passive particle and that of the active bath divided by $F_0^2$, as a function of $F_0$. The dotted black line represents the kinetic temperature of an isolated ALP.
\label{fig:temp_thermo}
}
\end{figure}

From our simulation data, we then determined the kinetic temperature of the thermostatted probe as defined by $k_\mathrm{B}T_\mathrm{eff}=m\langle\mathbf{V}^2(t)\rangle/d$. The results are shown in Fig.~\ref{fig:temp_thermo}.

\section{Second fluctuation dissipation theorem}
\label{sec:1fdt_supp}

\begin{figure}
  \centering
  \includegraphics[width=.9\linewidth]{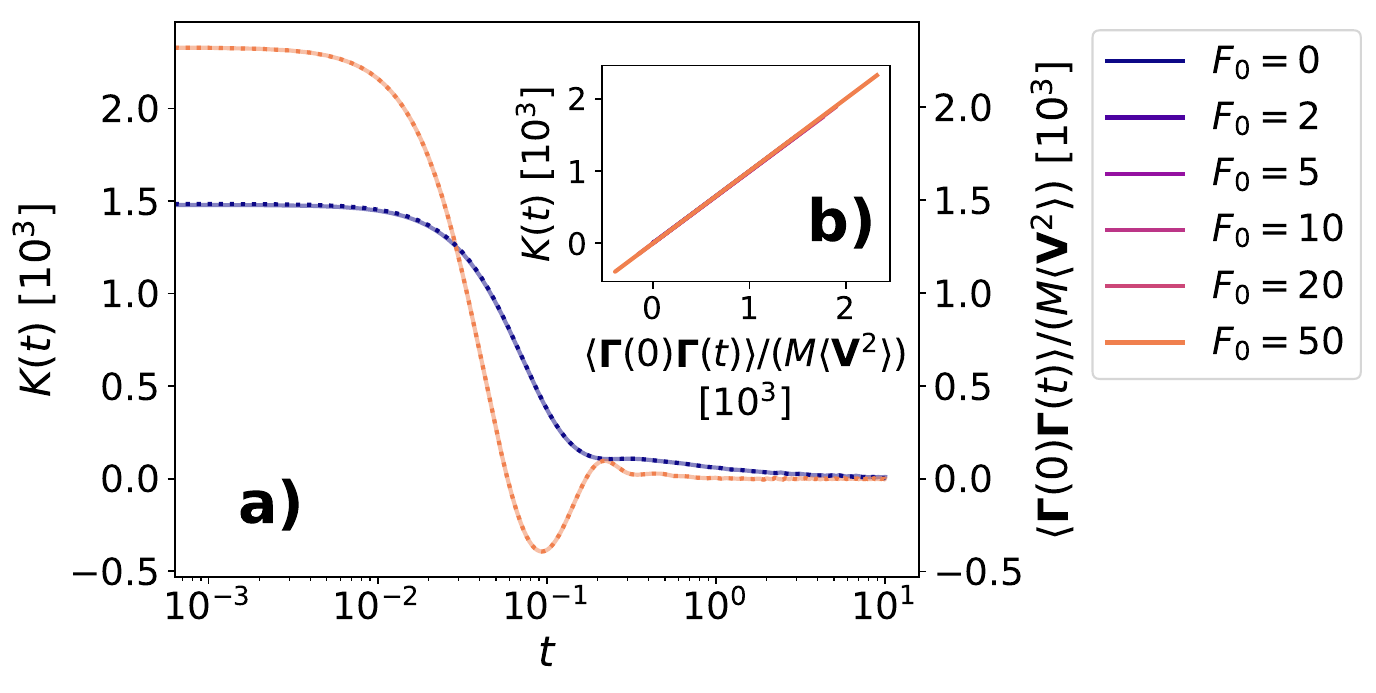}  
\caption{Test of the $2^{nd}$FDT for the immersed passive particle in a bath with density $\rho=0.4$. \textbf{a)} Memory kernel (solid line) and the stochastic force autocorrelation function (dashed line) as a function of time for values of $F_0=0$ and $F_0=50$. The prefactor of the memory kernel has been factored out. \textbf{b)} The stochastic force autocorrelation function as a function of the memory kernel (prefactor removed) for various $F_0$.}
\label{fig:fdt_conf}
\end{figure}

To fulfill the $2^{nd}$FDT, we expect the memory kernel, $K(t)$, to be identical, up to a prefactor of $M\langle |\mathbf{V}|^2\rangle$, to the autocorrelation function of the stochastic force. We define $K(t)$ through the Volterra equation:
\begin{equation}
\label{eq:iv}
m\dot{C}_{V}(t)=-\int^t_0\mathrm{d}s K(t-s) C_{V}(s),
\end{equation}
where $C_{V}(t)$ is the velocity autocorrelation function. While there have been many techniques created to extract the memory kernel from simulations, we use the technique proposed by Shin et al.~[34] because this is the most direct method of extraction and follows directly from solving Eq.~\eqref{eq:iv}. From a trivial rewriting of Eq.~(\ref{eq:gle}), we calculate the stochastic force on a particle directly from our simulation data~[34] and then calculate its autocorrelation function, $\langle \bm{\Gamma}(0) \bm{\Gamma}(t) \rangle$. We confirm the fulfillment of the $2^{nd}$FDT for the immersed probe in Fig.~\ref{fig:fdt_conf}.

\section{Entropy production}
\label{sec:entropy}
\begin{figure}
  \centering
  \includegraphics[width=1\linewidth]{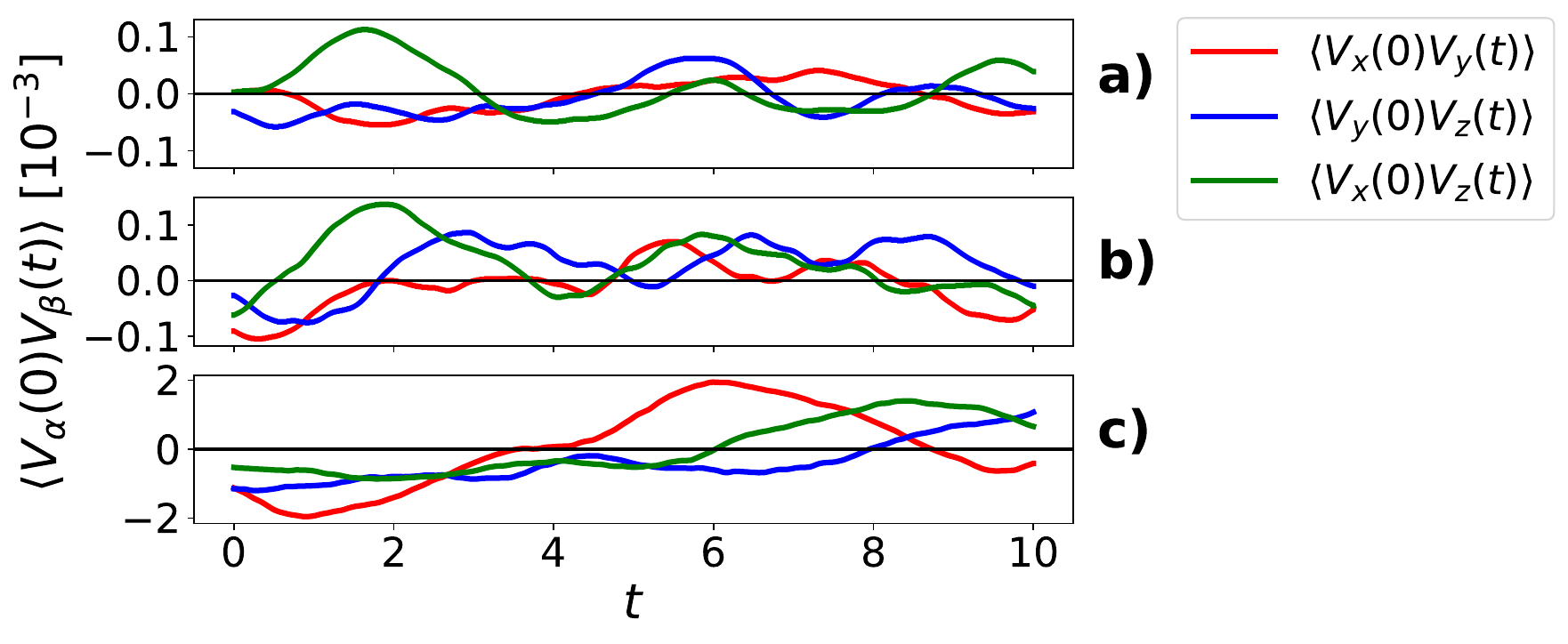}  
\caption{Velocity cross correlation functions of the probe particle in a fluid of density $\rho=0.4$ for activities \textbf{a)} $F_0=0$, \textbf{b)} $F_0=10$, \textbf{c)} and $F_0=50$. }
\label{fig:vcross}
\end{figure}

As a means of investigating the possibility of entropy production through the breaking of time reversal symmetry, we examine the stochastic force on the probe as well as the velocity cross-correlation functions of the probe. We find that the velocity cross-correlation function of the probe are zero, even in the presence of an active bath, as is shown in Fig.~\ref{fig:vcross}. Thus, we do not see any signatures of time reversal symmetry breaking in the velocity of the probe.

\begin{figure}
  \centering
  \includegraphics[width=.9\linewidth]{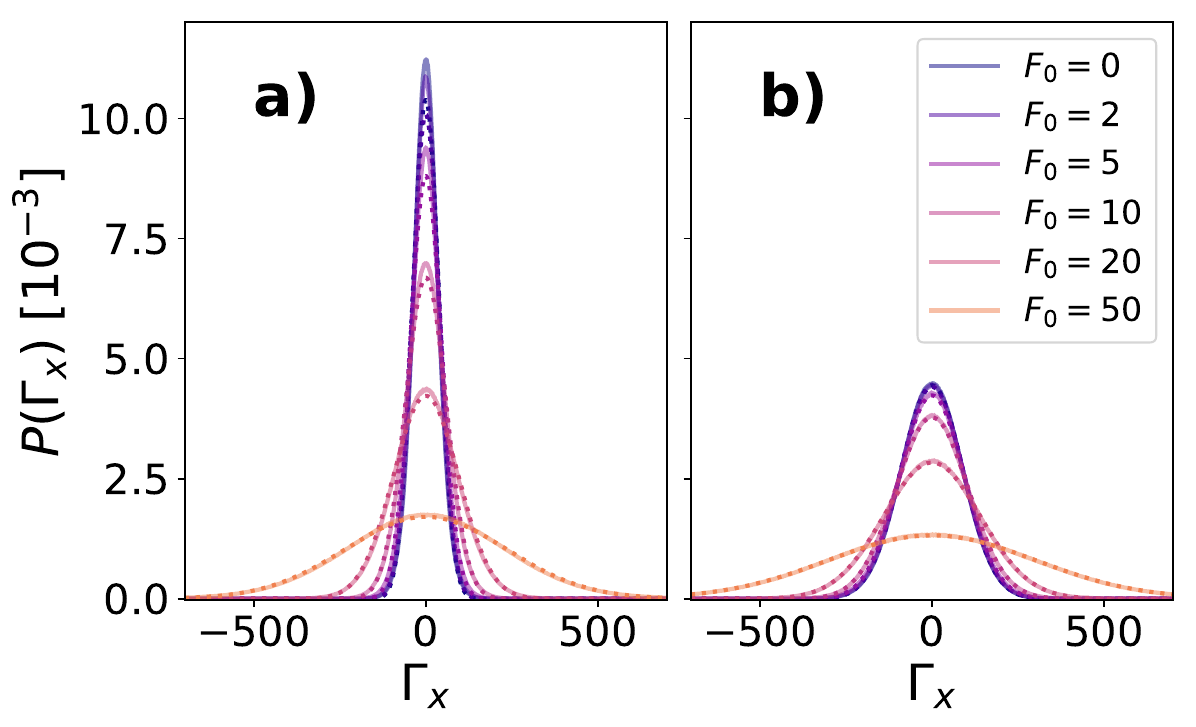}  
\caption{Stochastic force distribution on the probe particle for a fluid of ALPs with density \textbf{a)} $\rho=0.4$ and \textbf{b)} $\rho=0.8$.}
\label{fig:sf}
\end{figure}

From a trivial rewriting of Eq.~(\ref{eq:gle}), we calculate the stochastic force on the probe particle directly from our simulation data~[34]. We find that, in spite of the non-equilibrium nature of the bath, the stochastic force distribution on the probe remains Gaussian, as is shown in Fig.~\ref{fig:sf}. Although we do see some deviations from a Gaussian stochastic force distribution at low densities (see Fig.~\ref{fig:sf}a)), these deviations seem to be due to low density effects rather than to the bath activity. In fact, as the activity of the bath increases, the deviations from a Gaussian distribution at low densities actually decrease. Therefore, the stochastic force on the probe also does not reveal any signs of time reversal symmetry breaking.

\section*{Acknowledgements}
We wish to thank Thomas Speck and Ashreya Jayaram for useful discussions. This work was funded by the Deutsche Forschungsgemeinschaft (DFG) via Grant 233630050, TRR 146, Project A3. Computations were carried out on the Mogon Computing
Cluster at ZDV Mainz.

\bibliography{refs} 

\end{document}